\documentstyle[preprint,aps]{revtex}

\begin{document}
\input{epsf}
\draft
\title{THE EFFECT OF ELECTRON-ELECTRON INTERACTIONS ON THE
AVERAGE POLARIZABILITY OF A MESOSCOPIC SYSTEM}

\author
{Richard Berkovits}

\address{
The Jack and Pearl Resnick Institute of Advanced Technology,\\
Department of Physics, Bar-Ilan University,
Ramat-Gan 52900, Israel}

\date{\today}
\maketitle

\begin{abstract}
A comparison between an analytical calculation of the polarizability
of a mesoscopic interacting system in the random phase approximation and
numerical exact diagonalization results is presented. While
for weak interactions the analytical calculation fits the numerical results
rather well, deviations appear for stronger interactions. This is the
result of the appearance of intermediate range correlations in the electron
density, which suppresses the polarizability below its classical value. 
The relevance to quantum dot systems is discussed.
\end{abstract}
\pacs{PACS numbers: 71.55.Jv,73.20.Dx,71.27.+a}

The study of the polarizability of small mesoscopic systems has a long
history. Already in 1965 Gorkov and Eliashberg\cite{ge} 
have attempted to calculate
the average polarizability of a small metallic grain by applying the concept
of energy level repulsion from the random matrix theory (RMT)\cite{r1,r2}.
Their calculation, which can be backed by a non-linear sigma model 
calculation\cite{bm}, shows that the average polarizability
of a mesoscopic system $\langle \alpha \rangle \propto (\kappa L)^{(d-1)} 
L^3$ (here $L$ is the length of the system and the inverse
screening length $\kappa^{(d-1)} = S_d e^2 N(0)$,
where $N(0)$ is the level density at the Fermi energy, $S_d=2 \pi, (4 \pi)$
for two (three) dimensional systems and $\langle \ldots \rangle$ denotes 
an ensemble average). This result is much larger than
the classical value of the polarizability  $\alpha \propto L^3$.
The reason for this apparent enhancement of the polarizability is, as
pointed out by Strassler {\it et. al.} \cite{sa1}, the neglect of 
electron-electron (e-e) interactions which reduce the quantum
corrections dramatically. Once $\kappa L \gg 1$, which for metallic
systems corresponds to $L \gg 1 \AA$, the polarizability is proportional
to the volume\cite{sa2}. The effect of e-e interactions in the random phase
approximation (RPA) has been recently
incorporated into the RMT formulation of the polarizability
by Efetov \cite{efetov}. The fluctuations in the grand-canonical
ensemble were shown to be strongly suppressed by the e-e interactions
\cite{ba1,ba2}, while the suppression is less significant in the canonical
ensemble\cite{uz}.

In recent experiments on disordered quantum-dots\cite{rev}
it turned out that the 
ground-state energies of those dots show large fluctuations\cite{cow,sba}.
These fluctuations are larger than
expected from RMT theory in which the e-e interactions are incorporated
by the RPA\cite{sba}. This deviation from RMT theory is the result of the
appearance of intermediate range correlations in the electron density 
as a result of strong e-e interactions. In this paper we shall show that
the average polarizability is also affected by strong
e-e interactions. While in the weak interaction regime the average 
polarizability is proportional to $L^3$, for strong interactions
the polarizability tends to zero.

The polarizability may be written as
\begin{equation}
\alpha = e^2
{{\partial \ } \over {\partial E}} \int d \vec r (z - L/2)
n(r) \Phi(r),
\label{e1}
\end{equation}
where $n(r)$ is the electronic density, $\Phi(r)$ is the local electrostatic
potential as a result of applying an external electric field $E \hat z$ and
the polarizability is calculated in the $\hat z$ direction. 
We shall begin by considering the weak interaction limit. In this
limit one can treat both the disorder and the e-e interactions
perturbatively. Taking into account the e-e interactions in the RPA one may
rewrite Eq. (\ref{e1}) for the average polarizability as \cite{ba2}
\begin{equation}
\langle \alpha \rangle = e^2 N(0)
\int d \vec r \ d \vec r\prime (z - L/2) \chi(\vec r,\vec r\prime)
(z\prime - L/2),
\label{e2}
\end{equation}
which after assuming a rectangular geometry and performing a Fourier
transform results in
\begin{equation}
\langle \alpha \rangle = e^2 N(0) \bigg({{L}\over{2}}\bigg)^{2d}
\sum_{\vec q} [z(\vec q)]^2 \chi(\vec q),
\label{e3}
\end{equation}
where $\vec q= (n_x \pi / L) \hat x + (n_y \pi / L)  \hat y + 
(n_z \pi / L)  \hat z$
with $n_x, n_y, n_z =0,1,2 \ldots$ (of course for 2D systems
$\vec q= (n_x \pi / L)  \hat x + (n_z \pi / L)  \hat z$),
\begin{equation}
z(\vec q) = {{2^d}\over{L}} \delta_{q_x=0} \delta_{q_y=0} {(-1)^{(n_z+1)}
\over{q_z}^2},
\label{e4}
\end{equation}
and
\begin{equation}
\chi(\vec q) = \bigg({{2}\over{L}}\bigg)^d  {{1}\over{1+N(0) {\cal V} U}},
\label{e6}
\end{equation}
for short range interactions represented by an interaction potential
$U(\vec r, \vec r \prime) = {\cal V} U \delta (\vec r - \vec r \prime)$ 
(${\cal V} = a^d$ where $a$ is the range of
the interaction and $d$ is the systems dimensionality).
For the Coulomb
interaction $U(\vec r, \vec r \prime) = U a / |\vec r - \vec r \prime|$,
where $U$ is the strength of the Coulomb interaction between two particles at
distance $a$, and
\begin{equation}
\chi(\vec q) = \bigg({{2}\over{L}}\bigg)^d {{q^{(d-1)}}\over{q^{(d-1)} + 
S_d N(0) a U}}\ .
\label{e7}
\end{equation}
Inserting all the definitions in Eqs. (\ref{e4}-\ref{e7}) and performing
the summation in Eq. (\ref{e3}) results, for short range interactions, in
\begin{equation}
\langle \alpha \rangle = {{2^d}\over{S_d 90}} L^3
(\kappa L)^{(d-1)} 
\bigg(1+{{{\cal V} U} \over {L^d \Delta}}\bigg)^{-1},
\label{e8}
\end{equation}
where $N(0)=(L^d \Delta)^{-1}$, and
$\Delta$ is the single electron level spacing.
For long range interactions
\begin{equation}
\langle \alpha \rangle = {{2^d}\over{S_d 90}} L^3 
(\kappa L)^{(d-1)} 
\sum_{n=1} \bigg( n^{(5-d)}\left(n^{(d-1)} + {{S_d a U}\over{\pi L \Delta}}
\right) \bigg)^{-1}.
\label{e9}
\end{equation}
Note that by using the constant $U$ to describe the strength of interactions
between the electrons while using $e$ to describe the interaction between
the electric field and the electrons we artificially create two different
interaction scales. Nevertheless, this distinction is useful because it enables
us to treat the interactions between the electrons by some appropriate
effective interaction (say, short range interactions) while the interaction
between the electric field and the electrons contributes a prefactor
$(\kappa L)^{(d-1)}$ to the polarizability. Only in this sense can one 
understand the original treatment of Gorkov and Eliashberg\cite{ge}
of the polarizability of non-interacting electrons. If we insert
$U=e^2/a$ in Eq. (\ref{e9}) we obtain
\begin{equation}
\langle \alpha \rangle = {{2^d}\over{S_d 90}} L^3 
(\kappa L)^{(d-1)} 
\sum_{n=1} \bigg( n^{(5-d)}\left(n^{(d-1)} + (\kappa L/\pi)^{(d-1)}
\right) \bigg)^{-1},
\label{e10}
\end{equation}
which for $\kappa L \gg 1$ gives the expected classical behavior
\begin{equation}
\langle \alpha \rangle \sim {{2^d} \pi^{(d-1)}\over{S_d 90}} L^3 .
\label{e11}
\end{equation}
Thus the RPA approximation corresponds to the classical result
$\alpha \propto L^3$.

In order to confirm the above results and to check their range of validity
we have performed a numerical calculation of the polarizability
for a system of interacting electrons on a 2D cylinder
of circumference $L_x$ and height $L_z$.
We used the following tight-binding Hamiltonian:
\begin{eqnarray}
H= \sum_{k,j} (\epsilon_{k,j} + (j-(L_z+1)/2)E) 
a_{k,j}^{\dag} a_{k,j} - V \sum_{k,j}
(a_{k,j+1}^{\dag} a_{k,j} + a_{k+1,j}^{\dag} a_{k,j} + h.c)
+ H_{int}
\label{hamil}
\end{eqnarray}
where  $a_{k,j}^{\dag}$
is the fermionic creation operator,
$\epsilon_{k,j}$ is the energy of a site ($k,j$), which is chosen 
randomly between $-W/2$ and $W/2$ with a uniform probability and $V$
is a constant hopping matrix element. We set the coupling the electrons
and the external electric
field $E$ to be unity (i.e., $e=1$). $H_{int}$ is the interaction part of
the Hamiltonian which for the short range interactions is given by:
\begin{equation}
H_{int} = U \sum_{\{k,j>l,p\}} a_{k,j}^{\dag} a_{k,j}
a_{l,p}^{\dag} a_{l,p},
\label{hamil1}
\end{equation}
where $\{\ldots\}$ denotes the nearest-neighbor
pairs of sites, and for the Coulomb interaction is equal to
\begin{equation}
H_{int} = U \sum_{k,j>l,p} {{(a_{k,j}^{\dag} a_{k,j} - K)
(a_{l,p}^{\dag} a_{l,p} - K)} \over 
{|\vec r_{k,j} - \vec r_{l,p}|/s}},
\label{hamil2}
\end{equation}
where $K$ is a positive background maintaining the charge neutrality
and $s$ is the lattice constant. 

For a sample of $M$ sites and $N$ electrons,
the number of eigenvectors spanning the many body Hilbert space
is $m = (_N^M)$. The many-body Hamiltonian may be represented by an
$m \times m$ matrix which is numerically diagonalized and the ground state 
eigenvector $\Psi^E(k,j)$
is obtained for different values of the e-e interaction
and electric field $E$.
Here we consider a $4 \times 4$ lattice (i.e., $M=16$ sites) and
$N=4,8$ electrons.
We chose $W=8V$ for which this system is in the metallic regime \cite{bav}
and average the results over $500$ realizations for each value
of interaction strength. 
The polarizability is calculated by
\begin{equation}
\alpha = {{(d(E) - d(0))} \over {E}},
\label{d1}
\end{equation}
where
\begin{equation}
d(E) = \sum_{k,j} |\Psi^E(k,j)|^2 (j-(L_z+1)/2).
\label{d2}
\end{equation}

\begin{figure}
\centerline{\epsfxsize = 6in \epsffile{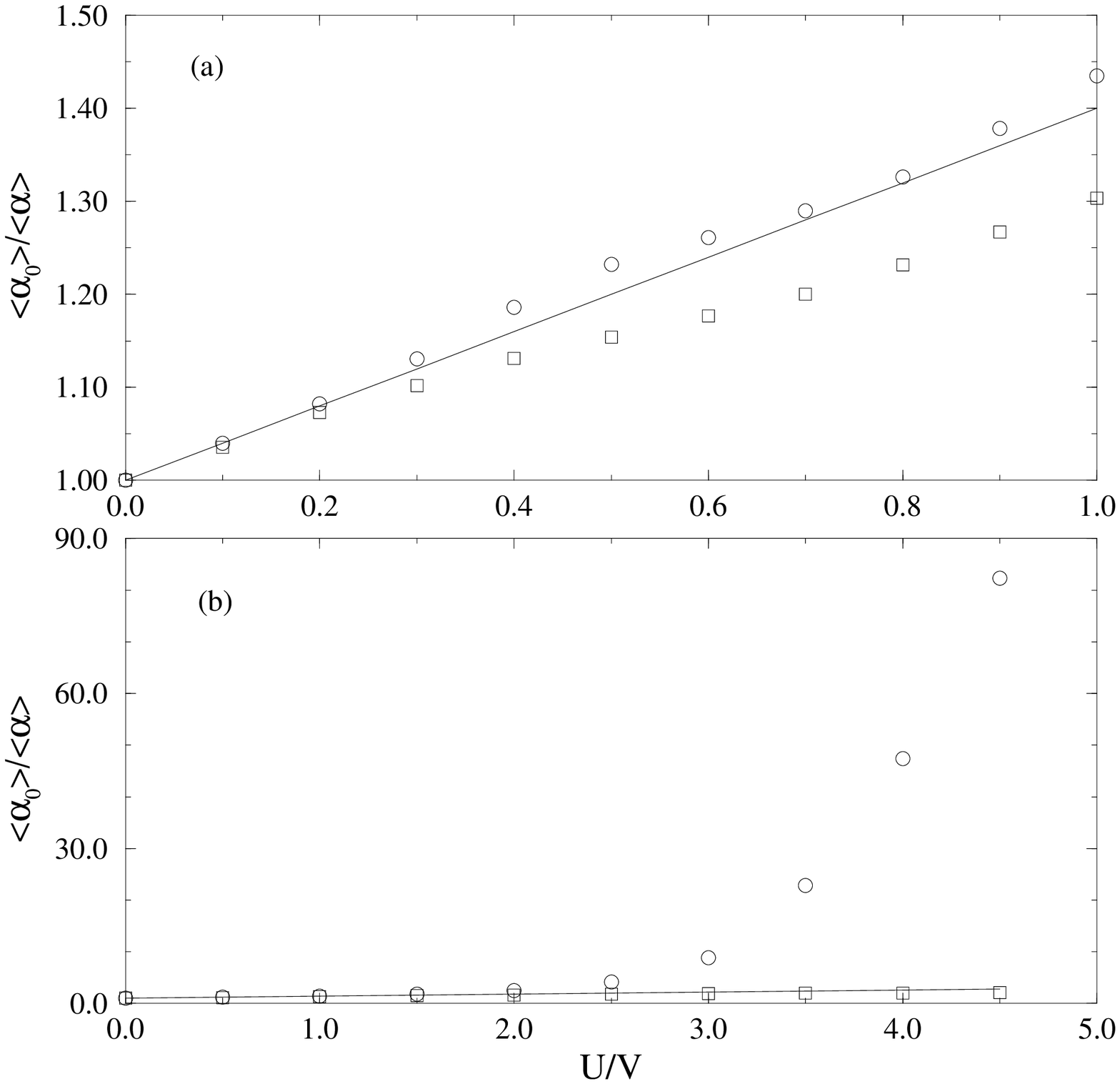}}
\caption {The ratio $\protect \langle \alpha(U=0) \rangle / 
\langle \alpha(U) \rangle$ for short range interactions. The numerical results
for $\protect N=8$ are represented by circles
and for $\protect N=4$ by squares.
The weak interaction regime is plotted in (a), while in (b) a larger range 
of interactions is presented.
\label{fig.1}}
\end{figure}

Our main goal is to study whether the effect of the e-e interactions is 
predicted correctly by Eqs. (\ref{e8}-\ref{e9}). Therefore it is convenient
to plot $\langle \alpha(U=0) \rangle / \langle \alpha(U) \rangle$ as function
of interaction strength,
which for short range interactions are is shown in
Fig. \ref{fig.1}. In order to evaluate Eq. (\ref{e8}) for the lattice
case one should replace ${\cal V} / L^d$ by $Z / (M-1)$ where the
average number of nearest neighbors, $Z$, in our case is 
given by $Z=3.5$ due to the
presence of boundaries. The average level spacing is $\Delta=0.58V$ resulting
in $\langle \alpha(U=0) \rangle / \langle \alpha(U) \rangle =
1 + 0.4 U$ plotted in the figure. We can see a reasonable correspondence
to the numerical values for both values of $N$
for weak interactions, i.e., $U<V$. For stronger values of interactions
strong deviations appear for $N=8$, while for $N=4$ the correspondence
holds. We shall discuss the reason for this behavior shortly.

\begin{figure}
\centerline{\epsfxsize = 6in \epsffile{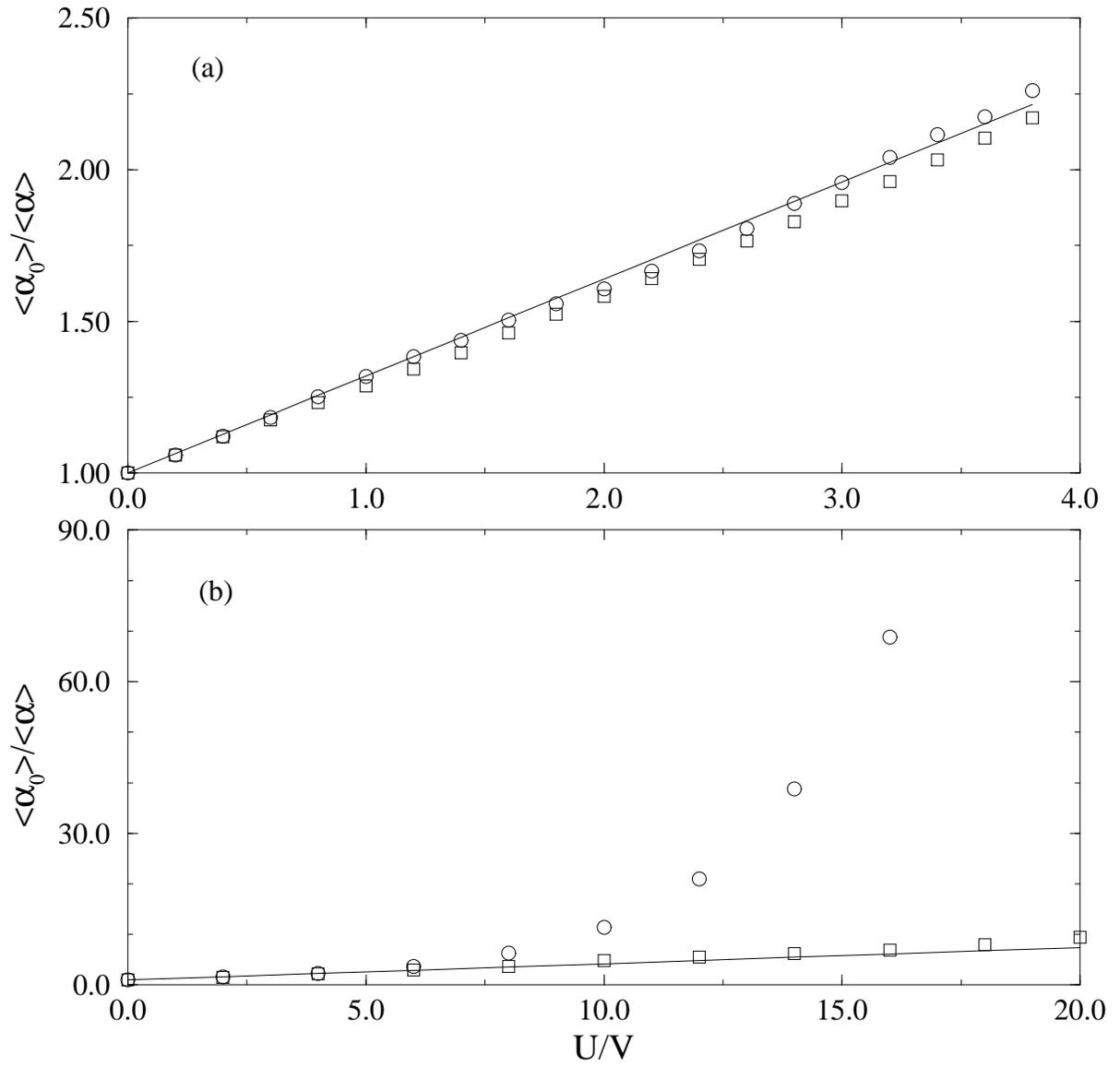}}
\caption {The same as in Fig. \protect{\ref{fig.1}} for the Coulomb
interactions.
\label{fig.2}}
\end{figure}

For the long range interactions the results are shown in Fig. \ref{fig.2}.
For a lattice system one should replace $S_2 = L^{-1} \int d^2r / r = 
2 \pi$ in Eq. (\ref{e9}) by ${S_2}^{*} = L^{-1} \sum_{k,j \ne l,p} 
|\vec r_{k,j} - \vec r_{l,p}|^{-1}$, which
in our case results in ${S_2}^{*} = 2.44$. Thus, after incorporating
${S_2}^{*}$ in Eq. (\ref{e9}) one obtains
$\langle \alpha(U=0) \rangle / \langle \alpha(U) \rangle =
1 + 0.32 U$ plotted in the figure. Again a good agreement between the theory
and the numerical results is seen for weak interactions ($U<4V$),
while deviations appear for both values of $N$ for strong interactions.

Thus, the RPA describes the polarizability rather well as long as the
interactions are not too strong, resulting in the classical polarizability
of Eq. (\ref{e11}) $\alpha \propto L^3$. Once the interactions are
strong the polarizability is suppressed compared to the classical value, i.e.,
$\alpha \ll L^3$.

\begin{figure}
\centerline{\epsfxsize = 6in \epsffile{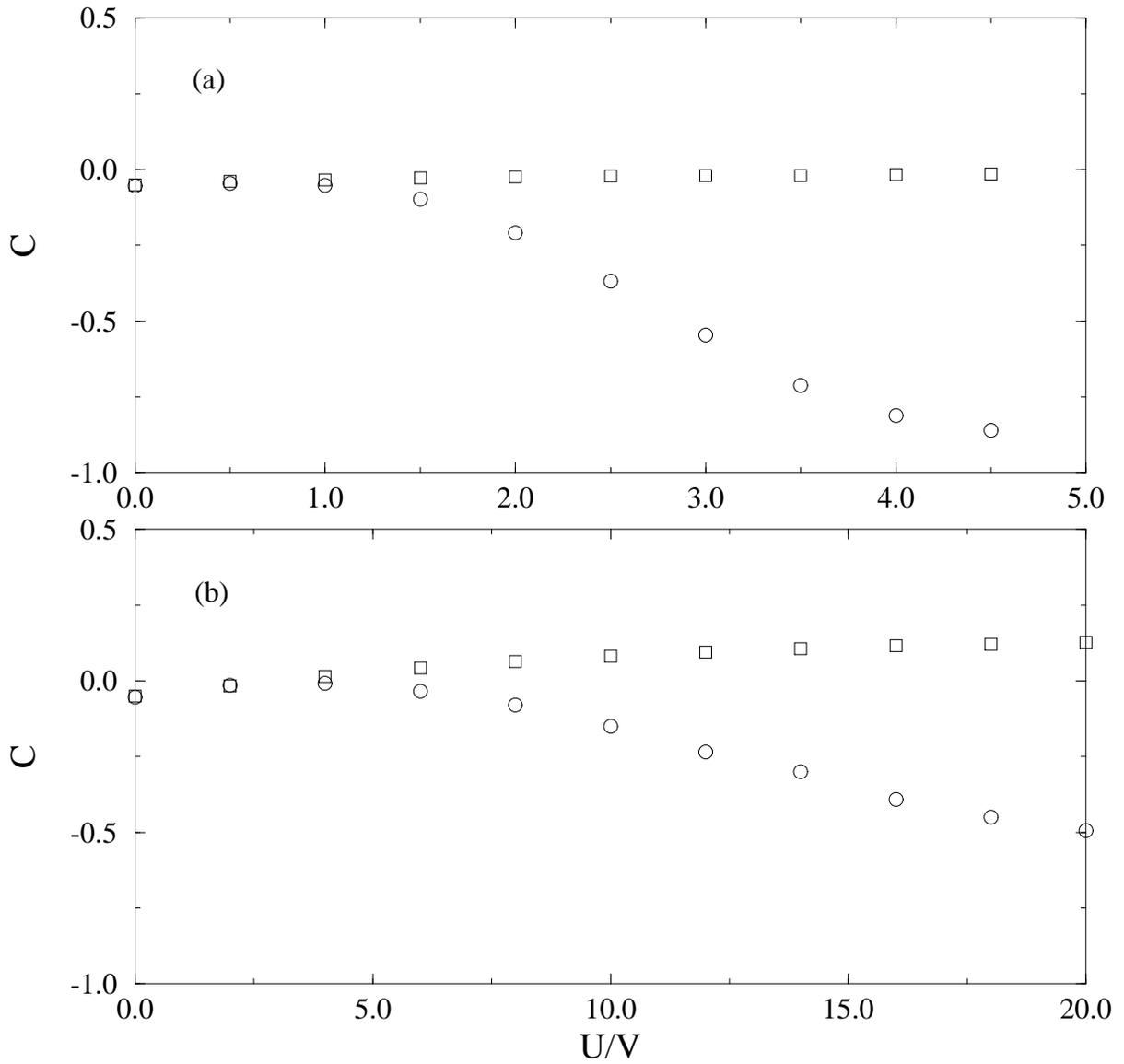}}
\caption {The two point  correlation $C(r=\protect {\sqrt{5}}s)$
for (a) short range
interactions, (b) long range interactions. Circles correspond to
$\protect N=8$ and squares to $\protect N=4$.
\label{fig.3}}
\end{figure}

The reason for the deviations is the appearance of intermediate range 
correlations for strong interactions. This may be clearly seen by defining 
a two point correlation function:
\begin{equation}
C(r) =  {{\sum_{k,j>l,p} C(\vec r_{k,j} - \vec r_{l,p} ) 
\delta_{|\vec r_{k,j} - \vec r_{l,p}|, r}} \over 
{\sum_{k,j>l,p} \delta_{|\vec r_{k,j} - \vec r_{l,p}|, r}}}\ ,
\label{corr}
\end{equation} 
where 
\begin{equation}
C(\vec r_{k,j} - \vec r_{l,p} ) = \left\langle
{{[|\Psi(k,j)|^2 - \langle |\Psi(k,j)|^2 \rangle]
[|\Psi(l,p)|^2 - \langle |\Psi(l,p)|^2  \rangle ]} 
\over 
{\langle |\Psi(k,j)|^2 \rangle\langle |\Psi(l,p)|^2\rangle}}
\right\rangle.
\label{corr1}
\end{equation}
In Fig. \ref{fig.3} we present the correlation $C(r=\sqrt{5}s)$
for different values of interaction. Under the RMT assumptions, as well as
for RPA, we expect $C(r=\sqrt{5}s) \rightarrow 0$. 
On the other hand, once Wigner crystallization
occurs we expect $C(r=\sqrt{5}s) \rightarrow -1$ for $N=8$, while
$C(r=\sqrt{5}s) \rightarrow 1$ for $N=4$. It can be clearly seen
that for weak interactions (i.e., $U<V$ for short range interactions and
$U<4V$ for Coulomb interactions) $C(r=\sqrt{5}s) \sim 0$. For stronger
interactions a transition towards the Wigner crystal values of $C(r=\sqrt{5}s)$
is clearly seen for both the Coulomb and the
short range interactions in the half filled case ($N=8$).
This correlation is a sign of the appearance of intermediate range order
in the electronic system. We call this intermediate range order and not long
range order since it decreases on the order of several lattice sites. Only
for stronger interactions does a real, long ranged, Wigner crystal appear.
Naturally, for lower fillings ($N=4$) the transition is much 
weaker for the Coulomb interactions and absent
for short range interactions. 

This behavior is the explanation for the deviations from the
perturbative results of the polarizability at strong interactions
seen in Figs. \ref{fig.1},\ref{fig.2}.
As long as there is no intermediate range order, RPA holds. 
Once intermediate range order appears, the
electrons are correlated, i.e., the electrons have a spatial order and
exhibit a strong resistance to being compressed. Therefore, the application of
an external electric field has less influence on their spatial distribution
and $\langle \alpha(U>>V) \rangle \rightarrow 0$ resulting in the strong
deviation of
$\langle \alpha(U=0) \rangle / \langle \alpha(U>>V) \rangle$ from the linear 
dependence especially
for the half-filled cases in which the effect of intermediate 
range order is the largest. 
For $N=4$ the deviations towards
larger values of $\langle \alpha(U=0) \rangle / \langle \alpha(U>>V) \rangle$
are weaker for the Coulomb interactions since the spatial correlations
are smaller and do not appear for the 
short range interactions where no correlations exist.

In order to consider the relevance of the above results to an experimental
situation one must determine the ration $U/V$. For quantum dots
$U/V > 3$ is realized, and strong evidence of the intermediate range order
appears in the behavior of the ground state energy fluctuations\cite{sba}. 
Therefore, one might expect that these systems will show a strong
decrease in their polarizability beyond the classical value
due to the appearance of intermediate range order.

I would like to thank A. Auerbach, B. L. Altshuler and U. Sivan for useful
discussions. I would also to acknowledge
the Minerva Center for the Physics of Mesoscopics, 
Fractals and Neural Networks and  the 
US-Israel Binational Science Foundation for financial support.

\end{document}